\documentclass{article}
\usepackage{lipsum}
\usepackage{spconf}

\usepackage[numbers, square, comma, sort&compress]{natbib}

\usepackage{graphicx}
\graphicspath{{figures/}}
\usepackage[labelformat=simple]{subcaption}

\DeclareCaptionLabelSeparator{periodspace}{.\quad}
\usepackage{array}
\usepackage{booktabs}
\newlength{\tabcoldefault}
\usepackage{url}

\usepackage{amsmath}
\usepackage{amssymb,mathrsfs,amsfonts}
\usepackage{mathtools}
\usepackage{icomma}
\usepackage[thmmarks, amsmath, standard, hyperref]{ntheorem}
\theoremstyle{nonumberplain}
\theoremseparator{.}
\renewtheorem{theorem}{Theorem}
\renewtheorem{proposition}{Proposition}
\renewtheorem{definition}{Definition}
\newtheorem{note}{Note}

\newcommand{\vect}[1]{{\mathbf{#1}}}
\newcommand{\mat}[1]{{\mathbf{#1}}}

\newcommand{\norm}[2]{\left\| #1 \right\|_{#2}}
\newcommand{\transpose}[1]{#1^\mathrm{T}}
\newcommand{\Real}{{\mathbb R}}

\newcommand{\vol}[1]{\operatorname{vol} \left( #1 \right)}

\usepackage[capitalise]{cleveref}
\crefname{section}{Section}{sections}
\crefname{figure}{Fig.}{figs.}
\crefname{equation}{Equation}{equations}
\crefrangelabelformat{equation}{(#3#1#4--#5#2#6)}
\crefname{problem}{Problem}{problems}
\creflabelformat{problem}{(#2#1#3)}
\crefname{algorithm}{Algorithm}{algorithms}
\creflabelformat{algorithm}{(#2#1#3)}
\crefname{assumption}{Assumption}{assumptions}
\creflabelformat{assumption}{(#2#1#3)}
\crefname{step}{Step}{steps}
\creflabelformat{step}{(#2#1#3)}
\crefname{algoline}{Line}{lines}

\usepackage{tikz}

\usepackage{flushend}

\title{A short-graph Fourier transform via personalized PageRank vectors}
\name{Mariano Tepper and Guillermo Sapiro\thanks{Work partially supported by NSF, ONR, NGA, ARO, and NSSEFF.}}
\address{Department of Electrical and Computer Engineering, Duke University}

\begin{document}

\maketitle

\begin{abstract}
    The short-time Fourier transform (STFT) is widely used to analyze the spectra of temporal signals that vary through time.
    Signals defined over graphs, due to their intrinsic complexity, exhibit large variations in their patterns.
    In this work we propose a new formulation for an STFT for signals defined over graphs. This formulation draws on recent ideas from spectral graph theory, using personalized PageRank vectors as its fundamental building block. Furthermore, this work establishes and explores the connection between local spectral graph theory and localized spectral analysis of graph signals.
    We accompany the presentation with synthetic and real-world examples, showing the suitability of the proposed approach.
\end{abstract}


\section{Introdution}

The Fourier transform globally decomposes a temporal signal into its constituting frequencies, identifying their contribution to the signal formation. Often, temporal signals vary their behavior through time; in these cases, the Fourier transform, being global, falls short as a tool to analyze the characteristics of these signals. The short-time Fourier transform (STFT)~\cite{Allen1977} is used to analyze the Fourier spectrum of temporally localized sections of the signal.
It is well studied that there is a trade-off between resolution (sharpness) in time  and its counterpart in frequency. There is no way to get arbitrarily sharp analysis in both domains simultaneously~\cite[Sec. 2.6.2]{Vetterli1995}.

Formally, in the STFT a window function $w$ which is nonzero for only a short period of time is slid along the time axis and multiplied by the input signal $f$; then the Fourier transform of the resulting signal is taken. Formally, for one dimensional signals,
\begin{equation}
\operatorname{STFT}_{f} (u, \xi) \doteq \int_{-\infty}^{\infty} f(t) w(t-u) e^{-i \xi t} dt .
\end{equation}
The changing spectra is usually analyzed as a function of the time-shift $u$ and is well suited to analyze time-varying signals. The above formula can be interpreted as the following three-sets algorithm: (1) translate the window $w$ by $u$, (2) modulate the result by frequency $\xi$, and (3) take the convolution of the result with the signal $f$.
This can be written as
\begin{equation}
\operatorname{STFT}_{f} = \langle f, \mathcal{M}_{\xi} T_u w \rangle ,
\end{equation}
where $(T_u g) (t) \doteq g(t - u)$ and
$(\mathcal{M}_{\xi} g) (t) \doteq g(t) e^{i \xi t}$ are the translation and modulation operators, respectively.

Weighted graphs are a natural representational structure in most modern network applications (including, for example, social, energy, transportation, and sensor networks). These graphs are loaded with information, usually in the form of high-dimensional data (i.e., signals) that reside on the vertices (nodes) of graphs. 
Graph signal processing lies at the intersection of graph theory and computational harmonic analysis and seeks to process such signals on graphs. See~\cite{Shuman2013,Sandryhaila2014} for further details and references on this emerging field.

Graph signal processing has been successful at characterizing the equivalent of the Fourier transform in graph domains.
Many different types of localized spectral transforms have been proposed in recent years, see~\cite[Sec.~IV]{Shuman2013} for a thorough discussion. This list includes a windowed Fourier transform~\cite{Shuman2012}, later described in this work.

One of the main examples are diffusion wavelets~\cite{Coifman2006}, which are based on compressed representations of powers of a graph diffusion operator.
In parallel, local spectral techniques, in which personalized PageRank vectors play a prominent role, have become increasingly popular in the field of community detection in graphs~\cite[e.g.,][]{Andersen2006a}.
As we will see in \cref{sec:sgft}, the PageRank equation is defined recursively, and we can consider a single PageRank vector in place of a sequence of random walk vectors, or of powers of a diffusion operator~\cite{Andersen2006}.

In this work, we establish and explore for the first time the connection between local spectral graph methods and localized spectral analysis of graph signals. This work is a first step in this exploration, and introduces a short-graph Fourier transform inspired on the ideas of local graph analysis, using personalized PageRank vectors as fundamental building blocks of the method.

%
%
%

The remainder of the paper is organized as follows.
In \cref{sec:sgft} we introduce our short-graph Fourier transform. Experimental results on synthetic and real graphs are presented in \cref{sec:experiments}, showing the interesting characteristics of the proposed formulation. Finally, we provide some concluding remarks in \cref{sec:conclusions}.


\section{From local spectral graph theory to a short-graph Fourier transform}
\label{sec:sgft}

We begin by introducing the notation and fundamental formulas used throughout the paper.

Let $\mat{X}$ be a matrix. In the following, $(\mat{X})_{ij}$, $(\mat{X})_{:j}$, $(\mat{X})_{i:}$ denote the $(i,j)$th entry of $\mat{X}$, the $j$th column of $\mat{X}$, and the $i$th row of $\mat{X}$, respectively.

We consider a graph $G = (V, \mat{A})$, where $|V|=n$ and $\mat{A} \in \Real_+^{n \times n}$ is the weighted adjacency matrix. The weighted entry $(\mat{A})_{ij}$ represents in most applications a measure of similarity between vertices $i$ and $j$. We assume that $G$ is connected and undirected, i.e., $(\mat{A})_{ij} = (\mat{A})_{ji}$.
The degree of a node $i \in V$ is $d_i \doteq \sum_{j = 1}^{n} (\mat{A})_{ij}$.
Let $\mat{D} \in \Real^{n \times n}$ be a diagonal matrix with entries $(\mat{D})_{ii} = d_i$.
The Laplacian of $G$ is defined as $\mat{L} \doteq \mat{D} - \mat{A}$, and the normalized Laplacian of $G$ is defined as $\mathcal{L} \doteq \mat{D}^{-1/2} \mat{L} \mat{D}^{-1/2}$.
We denote the eigendecompositions of $\mathcal{L}$ and $\mat{L}$ by $\mat{U} \mat{\Lambda} \transpose{\mat{U}}$ and $\mat{V} \mat{\Lambda}_{\mat{L}} \transpose{\mat{V}}$, respectively.
We assume that the eigenvalues, the diagonal entries of $\mat{\Lambda}$ and $\mat{\Lambda}_{\mat{L}}$, are sorted in increasing order.
Finally, the volume of a set of vertices $S \subseteq V$ is $\vol{S} \doteq \sum_{i \in S} d_i$.

%

Let $\vect{f} \in \Real^n$ be a signal over the graph vertices, i.e., $(\vect{f})_i$ is the signal value at vertex $i$.
The classical Fourier transform can be defined as the transform that diagonalizes the Laplace operator. Similarly, the graph Fourier transform~\cite{Chung1997} is defined as
$\widehat{\vect{f}} \doteq \transpose{\mat{V}} \vect{f}$,
where $\mat{V}$ diagonalizes the graph Laplacian. The inverse graph Fourier transform is then simply defined as
$\vect{f} \doteq \mat{V} \widehat{\vect{f}}$.

\subsection{Local spectral graph theory}

The second eigenvalue of the graph Laplacian $\mat{L}$ can be viewed as the solution to
\begin{equation}
    \min_{\vect{x}} \transpose{\vect{x}} \mat{L} \vect{x}
    \enspace \text{s.t.} \enspace
    \transpose{\vect{x}} \mat{D} \vect{x} = 1
    ,\enspace
    \transpose{\vect{x}} \mat{D} \vect{1} = 0 .
    \label[problem]{eq:spectral}
\end{equation}
The optimal solution $\vect{x}^*$ is a generalized eigenvector of $\mat{L}$ with respect to $\vect{D}$ and provides a map from the graph to the real line. This map encodes a measure of similarity (geodesic distance) between graph vertices. This property is exploited for clustering~\cite{normalizedCuts} and hashing~\cite{Weiss2008}, for example.

In~\cite{Mahoney2012} the above problem is modified to incorporate a bias towards a target region (defined by one or more vertices) in the graph. This region is represented as an indicator vector $\vect{s}$, normalized such that $\transpose{\vect{s}} \mat{D} \vect{1} = 0$ and $\transpose{\vect{s}} \mat{D} \vect{s} = 1$.
More precisely, given a set of nodes $S \in V$ we define the unit vector $\vect{s} = \operatorname{unit}(S)$ as
\begin{equation}
    (\operatorname{unit}(S))_i \doteq
    \begin{cases}
        b / \vol{S} \text{if } i \in S ;\\
        -b / \vol{V \smallsetminus S} \text{otherwise,}
    \end{cases}
\end{equation}
where $b = \sqrt{\vol{S} \vol{V \smallsetminus S} / \vol{V}}$.
The modified problem is given by~\cite{Mahoney2012}
\begin{equation}
    \min_{\vect{x}} \transpose{\vect{x}} \mat{L} \vect{x}
    \enspace \text{s.t.} \enspace
    \begin{gathered}
        \transpose{\vect{x}} \mat{D} \vect{x} = 1
        ,\enspace
        \transpose{\vect{x}} \mat{D} \vect{1} = 0
        ,\\
        \transpose{\vect{x}} \mat{D} \vect{s} \geq \kappa .
    \end{gathered}
    \label[problem]{eq:local_spectral}
\end{equation}
The only modification is the addition of the constraint on $\transpose{\vect{x}} \mat{D} \vect{s}$.
This can be interpreted as imposing to the solution a correlation with $\vect{s}$ larger than $\arccos(\kappa)$.

Intuitively, as the solution to \cref{eq:spectral} provides a notion of geodesic distance between graph nodes, the solution to \cref{eq:local_spectral} provides a notion of geodesic distance from the seed set to the rest of the vertices. This link will become clear in the following.

\begin{theorem}[\cite{Mahoney2012}]
    Let $\vect{s} \in \Real^n$ be a seed vector such that $\transpose{\vect{s}} \mat{D} \vect{1} = 0$ and $\transpose{\vect{s}} \mat{D} \vect{s} = 1$, and $\transpose{\vect{s}} \mat{D} \vect{v}_2 \neq 0$, where $\vect{v}_2$ is the second generalized eigenvector of $\mat{L}$ with respect to $\mat{D}$. In addition, let $\vect{x}^*$ be an optimal solution to \cref{eq:local_spectral} with correlation parameter $\kappa \in [0, 1]$. Then, there exists some $\gamma \in (-\infty, (\mat{\Lambda})_{2,2})$ and some $c \in [0, \infty]$ such that
    \begin{equation}
        \vect{x}^* = c\, \left( \mat{L} - \gamma \mat{D} \right)^+ \mat{D} \vect{s}.
    \end{equation}
    \label[theorem]{theo:local_spectral_equivalence}
\end{theorem}
PageRank~\cite{pagerank} assigns a numerical weight to each vertex of a graph, assessing its relative importance within the graph;
its personalized variant is frequently used to localize the PageRank vector within a subset of the network~\cite{Andersen2006}.
The following proposition can be proven using simple algebraic manipulations and the definition of $\mat{L}$.
\begin{proposition}
    Let $c = -\gamma = \tfrac{1 - \alpha}{\alpha}$ in \cref{theo:local_spectral_equivalence}.
    The vector $\vect{p}$, defined as
    $\vect{p} = c\, \left( \mat{L} - \gamma \mat{D} \right)^{-1} \mat{D} \vect{s}$,
    is the solution to the (degree normalized) personalized PageRank (PPR) equation
    \begin{equation}
    \mat{D} \vect{p} = (1 - \alpha) (\mat{D} \vect{s}) + \alpha \mat{A} \mat{D}^{-1} (\mat{D} \vect{p}) .
    \end{equation}
    \label{theo:ppr_link}
\end{proposition}

\cref{theo:local_spectral_equivalence,theo:ppr_link} connect \cref{eq:local_spectral} with the personalized PageRank equation~\cite{Jeh2003}. In the field of community detection, PPR vectors are used to find local communities around seed vertices~\cite[e.g.,][]{Andersen2006,Andersen2006a,Kloster2015}, where a small but cohesive ``seed set'' of vertices is expanded to generate its enclosing community (vertices having a stronger relationship to the seed set than to the rest of the graph). In this context, PPR vectors arise as natural units of observation for localized analysis of graphs.

Furthermore, powers of a graph diffusion operator were identified in~\cite{Coifman2006} as natural building blocks to define wavelets on graphs. Notice that the PPR vector is \emph{exactly} equivalent to the recursive application of the diffusion operator $\mat{A} \mat{D}^{-1}$~\cite{Andersen2006}, which leads naturally to the notion of geodesic distance.

Given this evidence, we posit that the PPR vector is a fundamental tool to perform a localized spectral analysis of graph signals. This connection is the key observation of this work and drives our definition of a short-graph Fourier transform.

\subsection{A short-graph Fourier transform}

As described in the introduction, we need two elements to define a short-graph Fourier transform: a localization (e.g., classically a translation) and a modulation operators.

\begin{definition}[Localization]
    We define the local window at node $i$ as 
    \begin{equation}
        \vect{w}_i \doteq \max \left(0,\ \vect{x}_i^* \right) / \norm{\max \left(0,\ \vect{x}_i^* \right)}{1} ,
    \end{equation}
    where $\vect{x}_i^*$ the solution to \cref{eq:local_spectral} with $\vect{s} = \operatorname{unit}(\{ i \})$ and the maximum is taken entrywise. 
    \label{def:sgft_localization}
\end{definition}
The window $\vect{w}_i$ is defined in terms of its correlation with $\operatorname{unit}(\{ i \})$, yielding to a simple conceptual interpretation.
In this work, we solve \cref{eq:local_spectral} using \cref{theo:local_spectral_equivalence}.
Given the eigendecomposition of the normalized Laplacian, we have $\mat{L} = \mat{D}^{1/2} \mat{U} \mat{\Lambda} \transpose{\mat{U}} \mat{D}^{1/2}$. Then,
\begin{equation}
    \vect{x}^* = c\, \left( \mat{D}^{-1/2} \mat{U} \right) \left( \mat{\Lambda} - \gamma \mat{I} \right)^+ \transpose{\left( \mat{D}^{-1/2} \mat{U} \right)} \mat{D} \vect{s}.
    \label{eq:sgft_localization}
\end{equation}
Once the eigendecomposition is computed as a preprocessing step, this formula delivers an efficient method for obtaining $\vect{x}^*$, without any iterations nor matrix inversions (albeit the inversion of the diagonal matrix $\mat{\Lambda} - \gamma \mat{I}$).
Interestingly, the spectral localization of $\vect{x}^*$ is determined by the product $\transpose{\left( \mat{D}^{-1/2} \mat{U} \right)} \mat{D} \vect{s} = \transpose{\mat{U}} \mat{D}^{1/2} \vect{s}$, i.e., by the correlation between $\vect{s}$ and each element of the graph Fourier basis.

Since the eigendecomposition of $\mathcal{L}$ is used in \cref{eq:sgft_localization}, we also use it in our graph modulation operator.
\begin{definition}[Graph modulation]
    For $k \in \{1, 2, \dots, n\}$, we define the graph modulation operator $M_k : \Real^n \rightarrow \Real^n$ by
    \begin{equation}
        M_k f \doteq \sqrt{\vol{V}} \ f \circ (\mat{D}^{-1/2} \mat{U})_{:k},
    \end{equation}
    where $\circ$ denotes the entrywise multiplication.
    \label{def:sgft_modulation}
\end{definition}
$M_1$ is the identity operator, such as $\mathcal{T}_0$ is in the classical modulation for temporal signals.
\begin{definition}
    Given the localization and modulation operators, we define the short-graph Fourier transform of a signal $\vect{f} \in \Real^n$ at vertex $i \in V$ and frequency $k \in \{ 1, 2, \dots, n \}$ as
    \begin{equation}
        \operatorname{SGFT}_{f}(i, k) = \langle \vect{f}, M_{k} \vect{w}_i \rangle .
        \label{eq:sgft}
    \end{equation}
    The spectrogram of $\vect{f}$ is defined as
    \begin{equation}
    \operatorname{spectrogram}_{f}(i, k) = | \operatorname{SGFT}_{f}(k, i) |^2 .
    \end{equation}
\end{definition}
It is not hard to see that \cref{eq:sgft} reduces to the standard one when the unweighted graph is a Cartesian grid.

\begin{note}
Shuman et al.~provide a different definition for a short-graph Fourier transform~\cite{Shuman2012}. They define the graph modulation operator as
$\widetilde{M}_k f \doteq \sqrt{n} \ f \circ (\mat{V})_{:k}$,
where $\circ$ denotes the entrywise multiplication. They also define a convolution operator as
$f \ast g \doteq \mat{V} \left( \widehat{f} \circ \widehat{g} \right)$,
and a translation operator as
$
    T_i f \doteq \sqrt{n} \ (f \ast \delta_i) =
    \sqrt{n} \ \mat{V} \left( \left(\transpose{\mat{V}} \right)_{:i} \circ \widehat{f} \right)
$,
where $\delta_i$ is the impulse function at vertex $i$.
This leads to the (more classical) definition 
$\operatorname{SGFT}_{f}(i, k) = \langle \vect{f}, \widetilde{M}_{k} T_i \vect{g} \rangle$.
Defining an appropriate window (kernel) $\vect{g}$ is not trivial in the graph space.
It is possible to define it in the graph spectral space as $(\widehat{\vect{g}})_{k} = e^{- \tau \lambda_k}$ and then invert the graph Fourier transform.
In this work we do not aim at producing a better method than the one in~\cite{Shuman2012} (although we will exemplify potential localization advantages of our definition). The method here proposed is based on radically different principles, which are of interest by themselves for the spectral study of graph signals.
\end{note}

\section{Experimental Results}
\label{sec:experiments}

We implemented the proposed short-graph Fourier transform in Python, using the graph-tool library~\cite{graph-tool}. We make the code publicly available at \url{https://github.com/marianotepper/sgft}. In all examples we set $\gamma = (\mat{\Lambda})_{1,1} - \beta$, where $\beta$ takes a particular value in each example.

\begin{figure}[t]
    \centering
    
    \begin{subfigure}{\columnwidth}
        \includegraphics[width=\columnwidth]{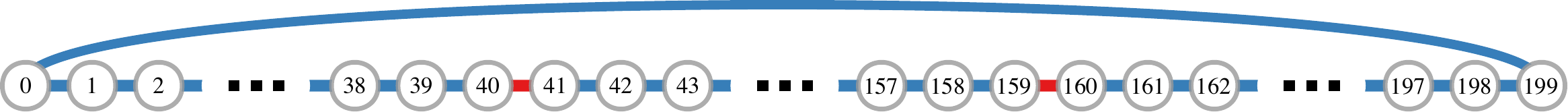}
        \caption{The linear graph has $200$ vertices, where each vertex is connected to its two neighbors (with periodicity in the edges). In the weighted case, the graph weights for all edges are set to one, excepting edges $(40, 41)$ and $(159, 160)$ (marked in red) which have a weight of $10^{-3}$.}
    \end{subfigure}
    
    \begin{subfigure}{\columnwidth}    
        \setlength{\tabcolsep}{0pt}
        \begin{footnotesize}
        \begin{tabular}{*4{c}}
            \multicolumn{2}{c}{Unweighted} & \multicolumn{2}{c}{Weighted} \\
            \cmidrule(lr){1-2} \cmidrule(lr){3-4}
            Conv. ($\tau = 200$) & PPR ($\beta=10^{-4}$) & Conv. ($\tau = 200$) & PPR ($\beta=10^{-4}$)\\
            \includegraphics[width=.25\columnwidth]{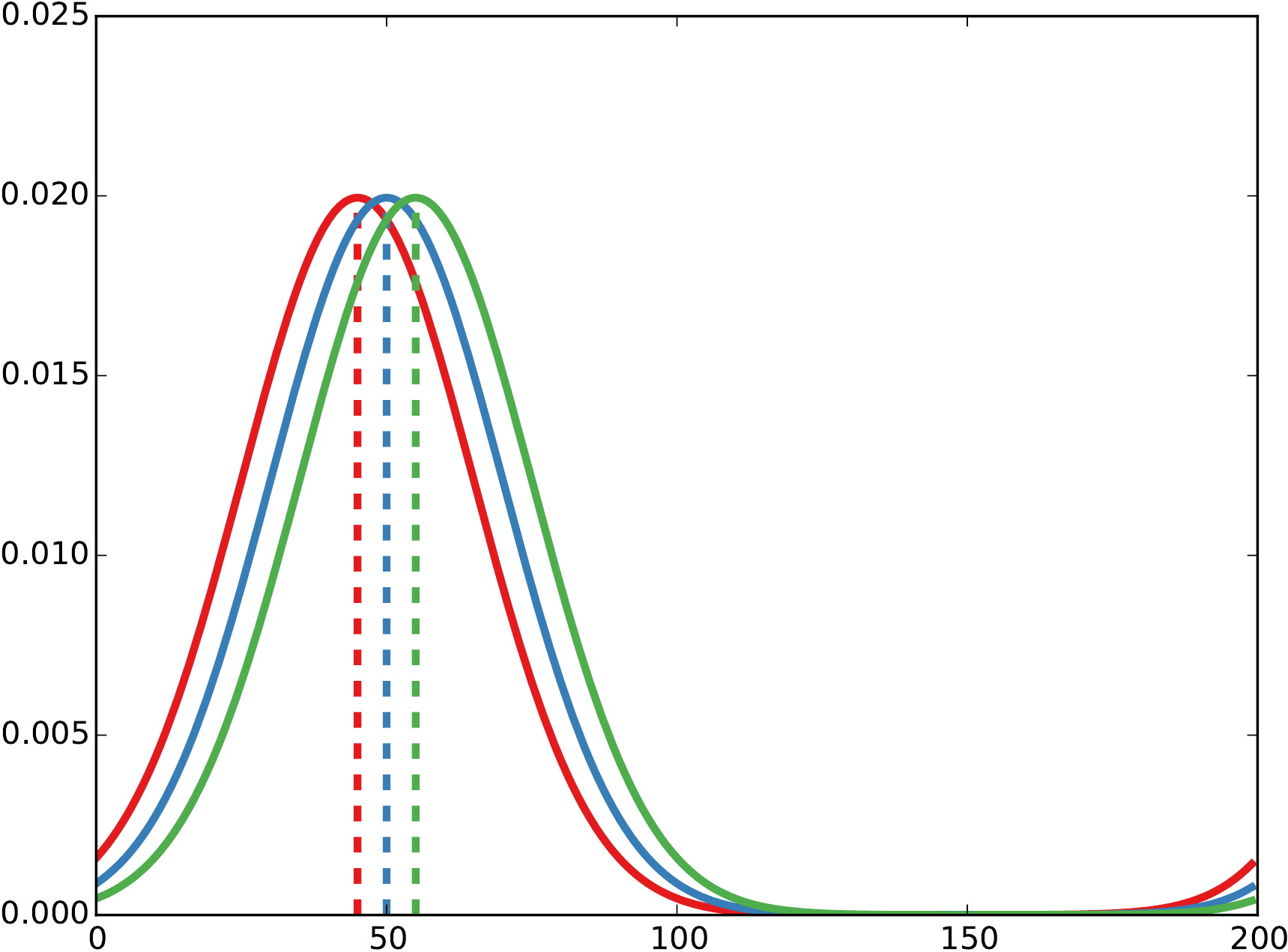}&
            \includegraphics[width=.25\columnwidth]{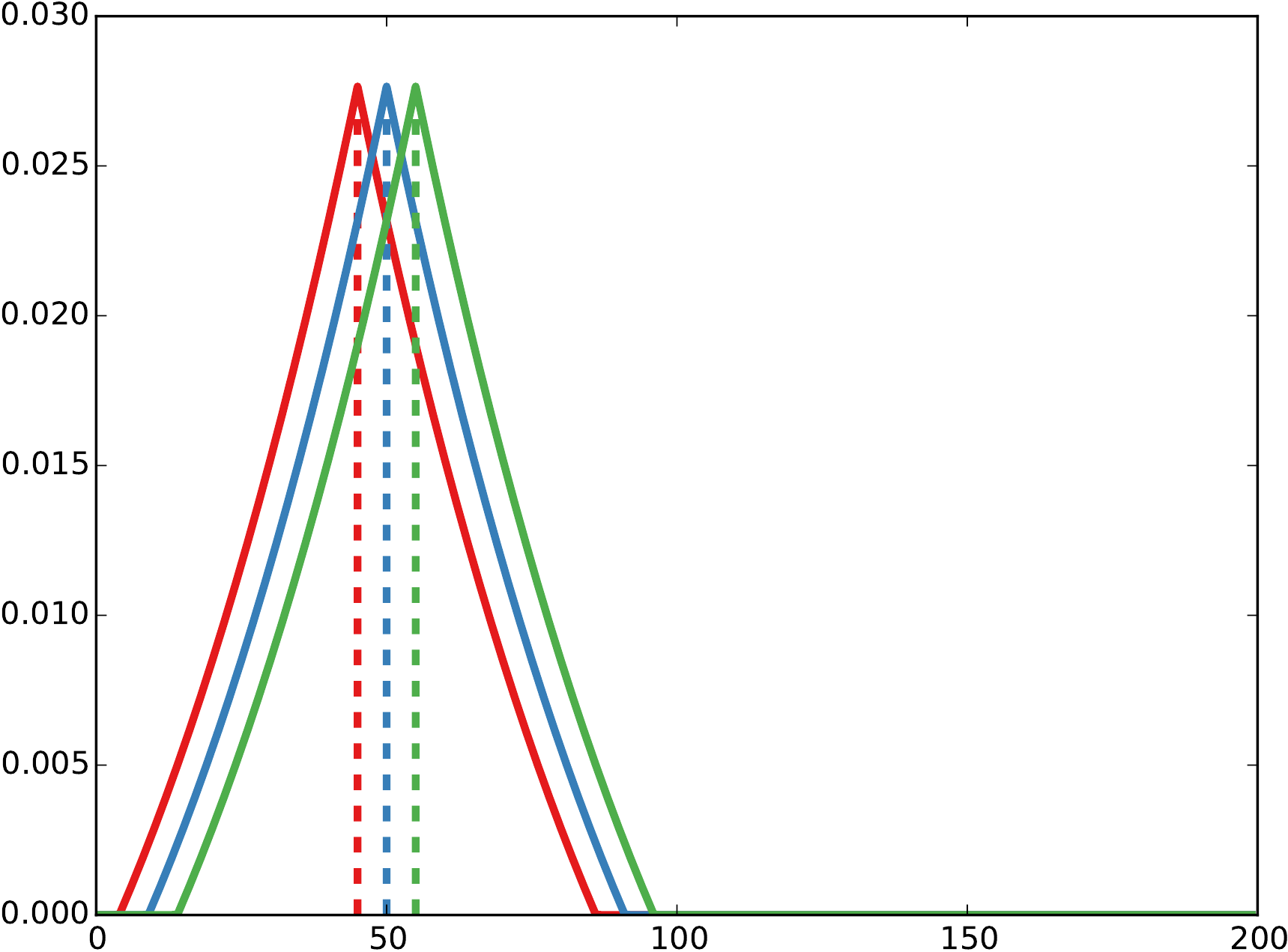}&
            \includegraphics[width=.25\columnwidth]{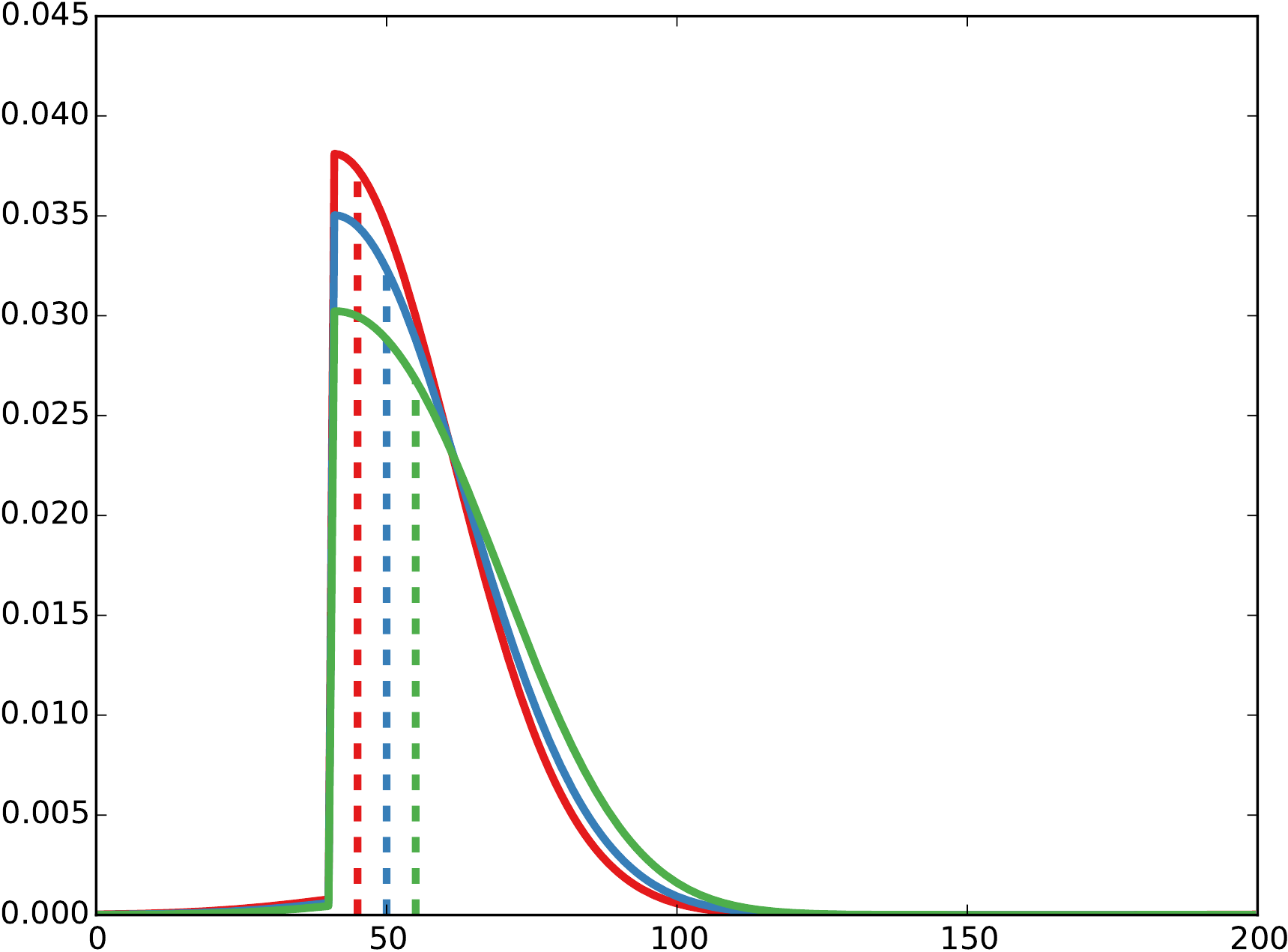}&
            \includegraphics[width=.25\columnwidth]{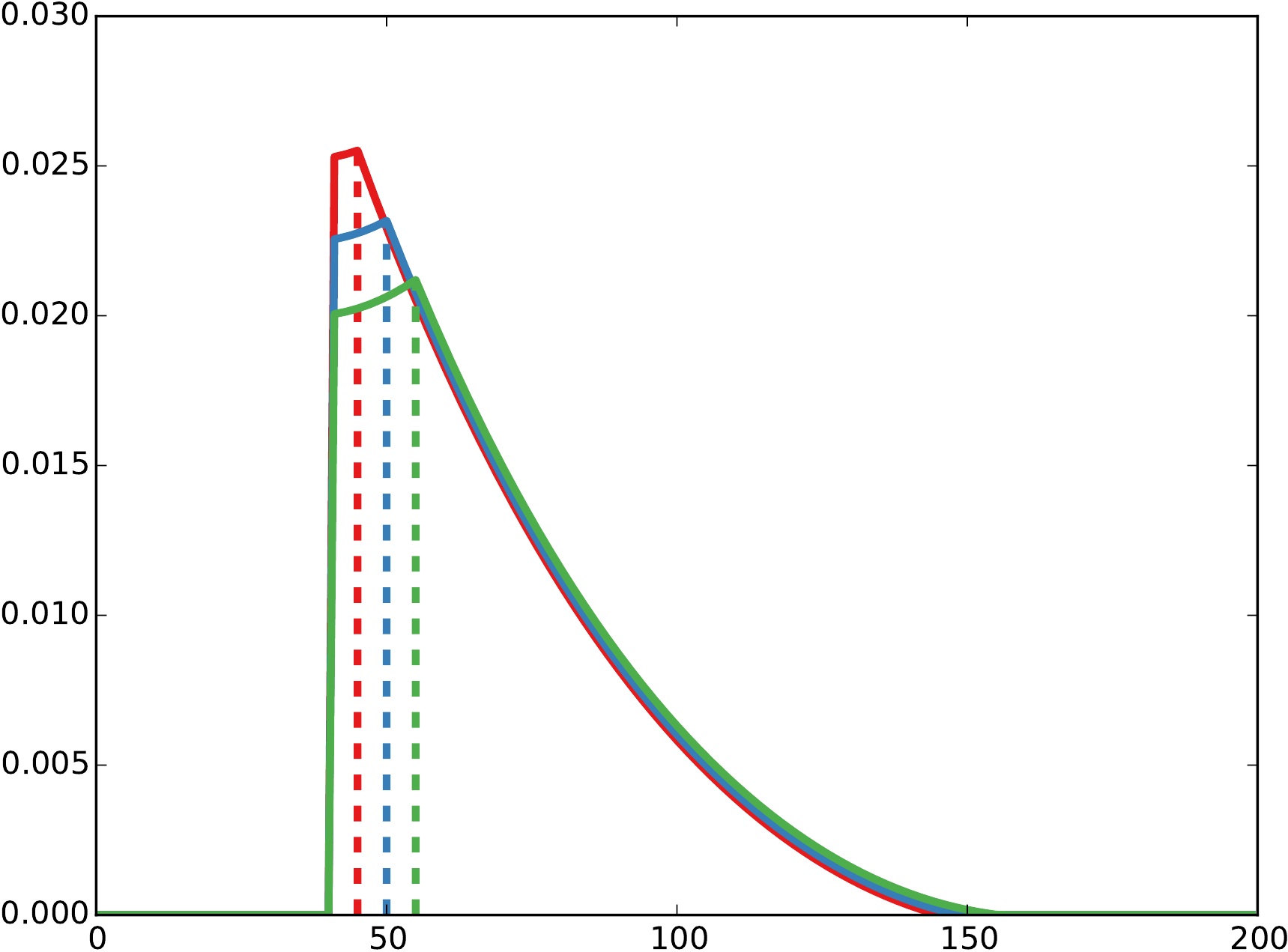}\\
        \end{tabular}
        \end{footnotesize}
        \setlength{\tabcolsep}{\tabcoldefault}
        
        \caption{We compute localized windows around vertices $45$, $50$, and $55$ (red, blue and green curves, respectively). Note how the convolutional approach~\cite{Shuman2012} fails to properly locate the window in the weighted case, as the window peak does not coincide with the desired vertex.}
    \end{subfigure}
    
    \caption{\textbf{Window localization comparison.} In the unweighted case (in which the graph Fourier and the standard Fourier transforms are equal), convolutional localization~\cite{Shuman2012} works well; however, when the graph weights present a sharp discontinuity, it fails to provide an accurate result. Contrarily, the proposed PPR approach works well in both cases.}
    \label{fig:localization}
\end{figure}

We begin by examining the localization operator (\cref{def:sgft_localization}). For this, we use in \cref{fig:localization} a linear graph, where localization can be easily interpreted and visualized. The main observation is that the window is properly localized when using the proposed approach, while this is not always the case with the convolutional approach.

\begin{figure}[t]
    \centering
    
    \parbox[c]{.27\linewidth}{%
        \includegraphics[width=.25\columnwidth]{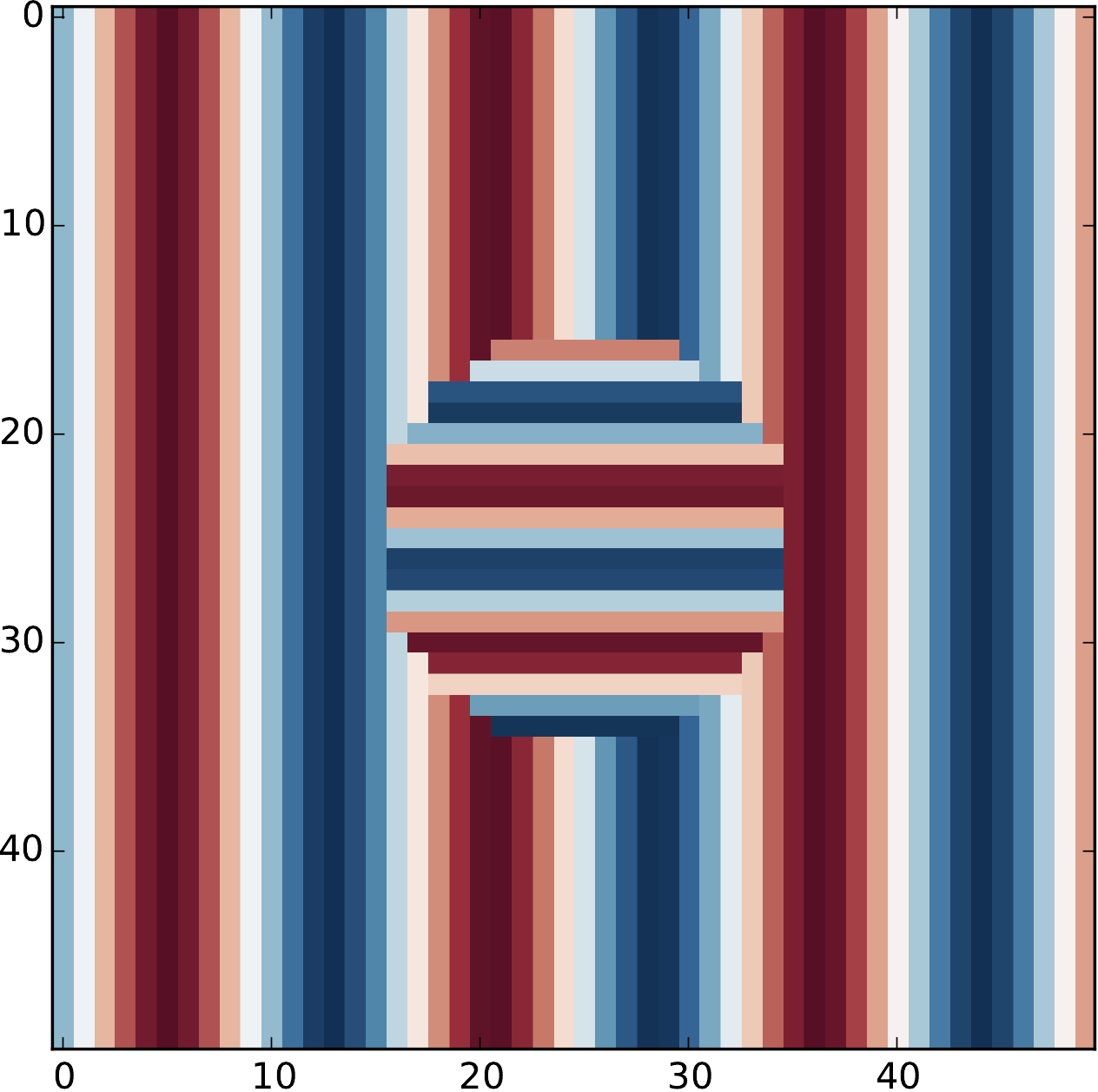}
    }
    \hfill
    \parbox[c]{.70\linewidth}{%
        \subcaption{The graph is a $50 \times 50$ regular grid, where each vertex is connected to its four neighbors (with periodicity in the edges). The input signal is formed by two sinusoidal waveforms, as shown on the side. In the weighted case, the graph weights for the edges connecting both waveforms are set to $10^{-5}$ while for the rest of the edges, they are set to one.}
    }
    
    \vskip.5em
    
    \begin{subfigure}{\columnwidth}
        \centering
        
%
%
        
        \setlength{\tabcolsep}{0pt}
        \begin{footnotesize}
            \begin{tabular}{*4{c}}
                \multicolumn{2}{c}{Unweighted} & \multicolumn{2}{c}{Weighted} \\
                \cmidrule(lr){1-2} \cmidrule(lr){3-4}
                Conv. ($\tau = 5$) & PPR ($\beta=10^{-4}$) & Conv. ($\tau = 200$) & PPR ($\beta=10^{-4}$)\\
                
                \includegraphics[width=.25\columnwidth]{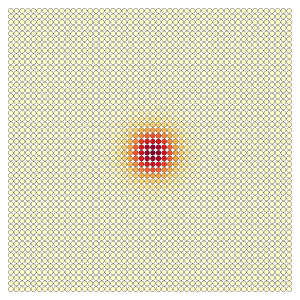}&
                \includegraphics[width=.25\columnwidth]{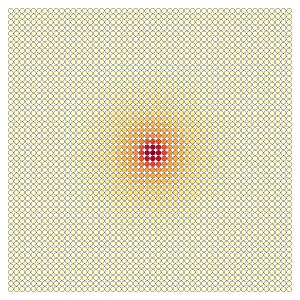}&
                \includegraphics[width=.25\columnwidth]{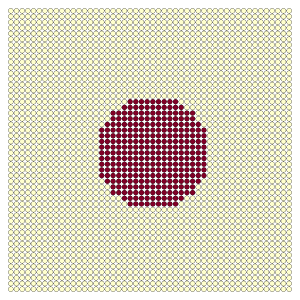}&
                \includegraphics[width=.25\columnwidth]{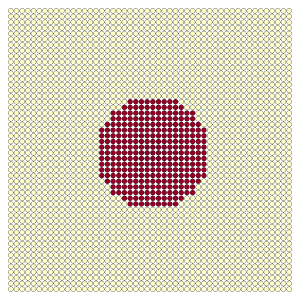}\\

                \includegraphics[width=.25\columnwidth]{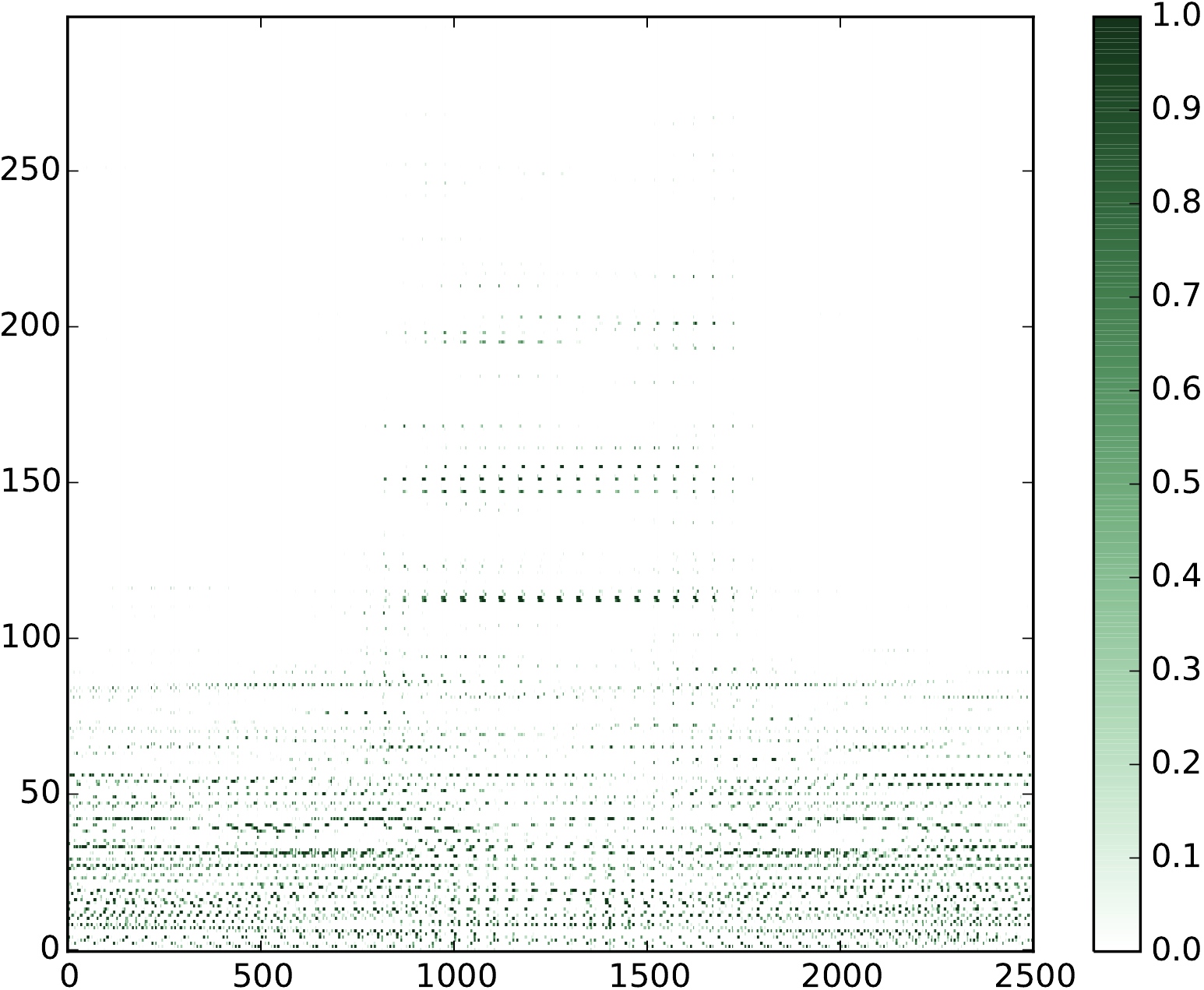}&
                \includegraphics[width=.25\columnwidth]{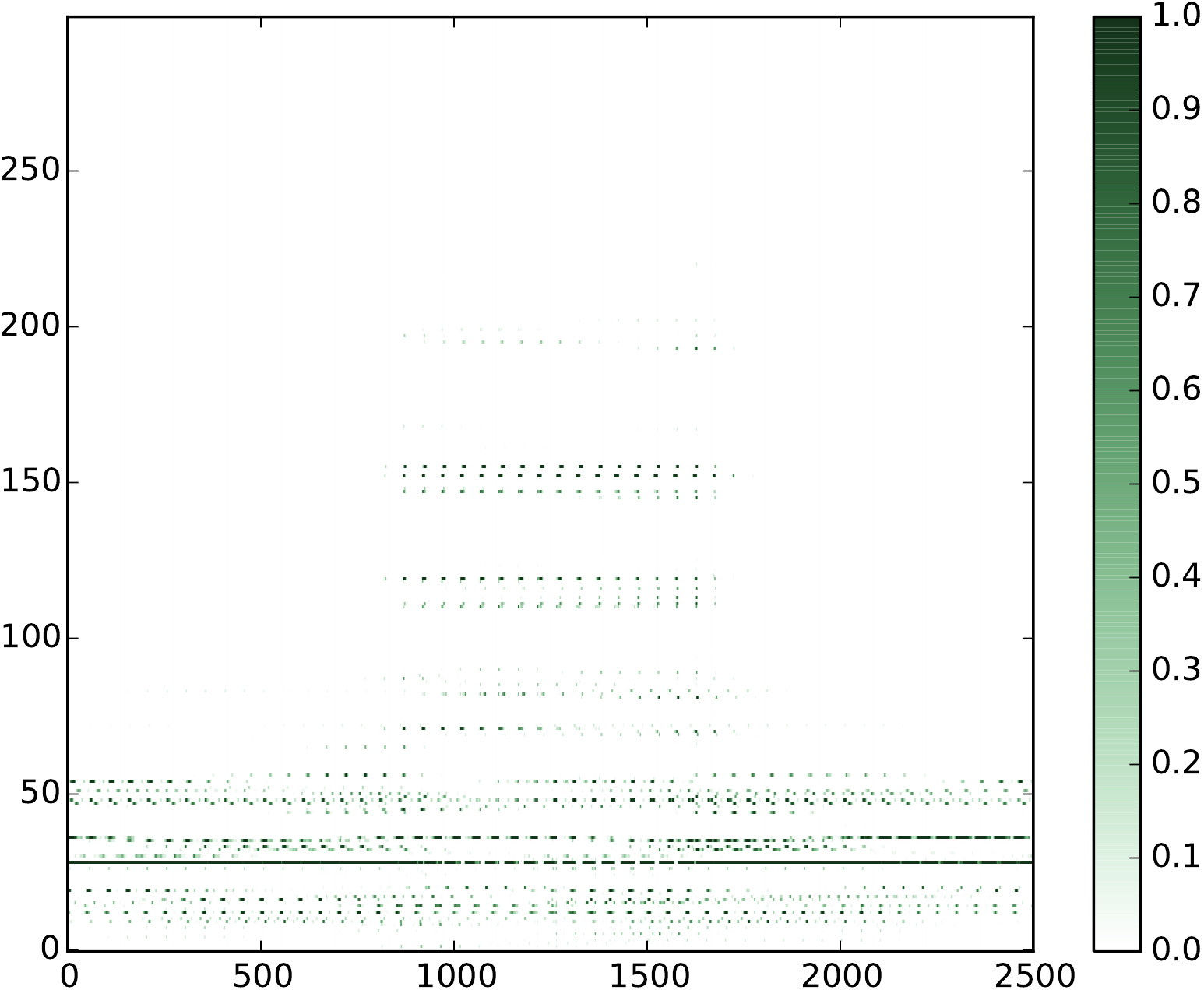}&
                \includegraphics[width=.25\columnwidth]{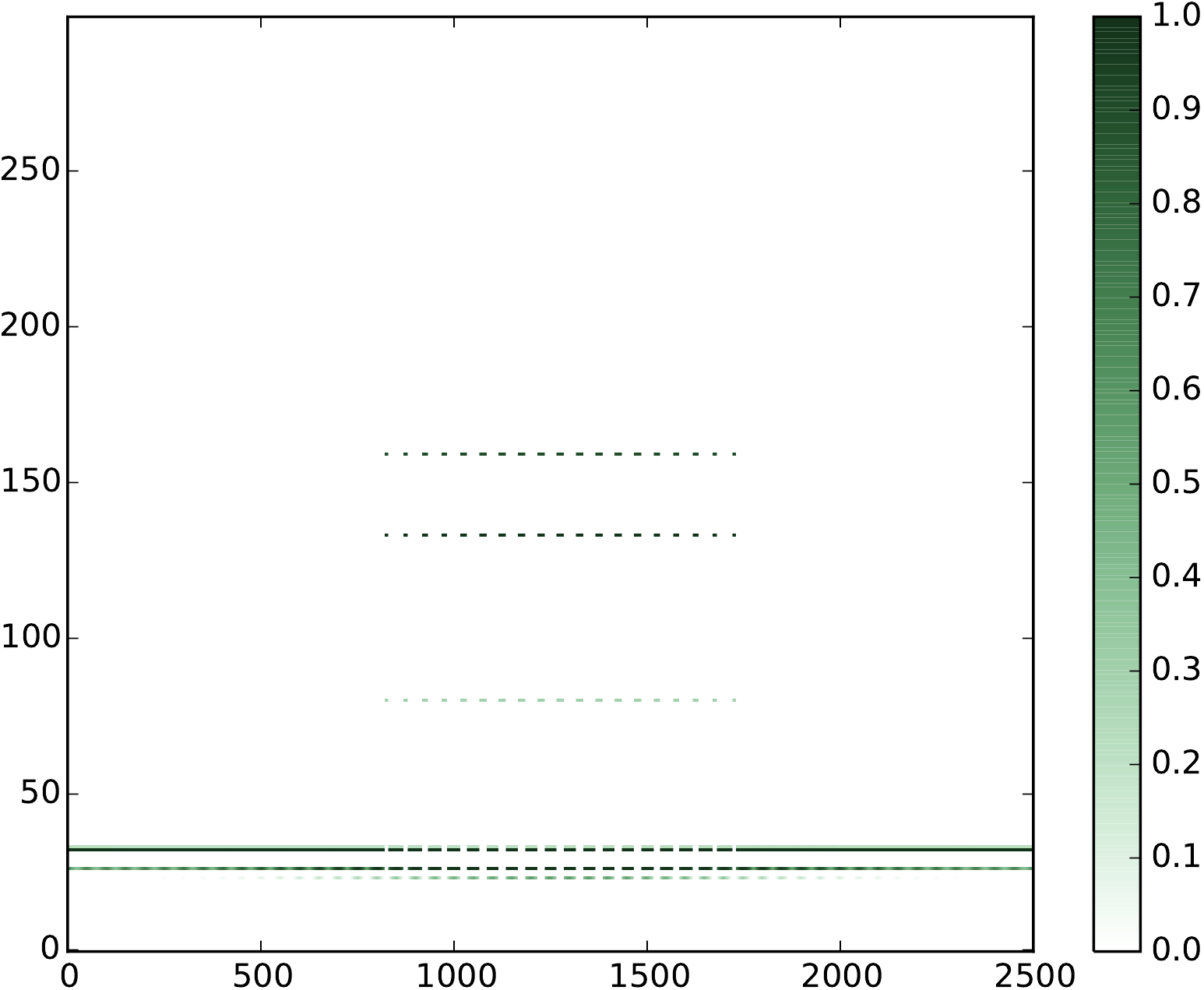}&
                \includegraphics[width=.25\columnwidth]{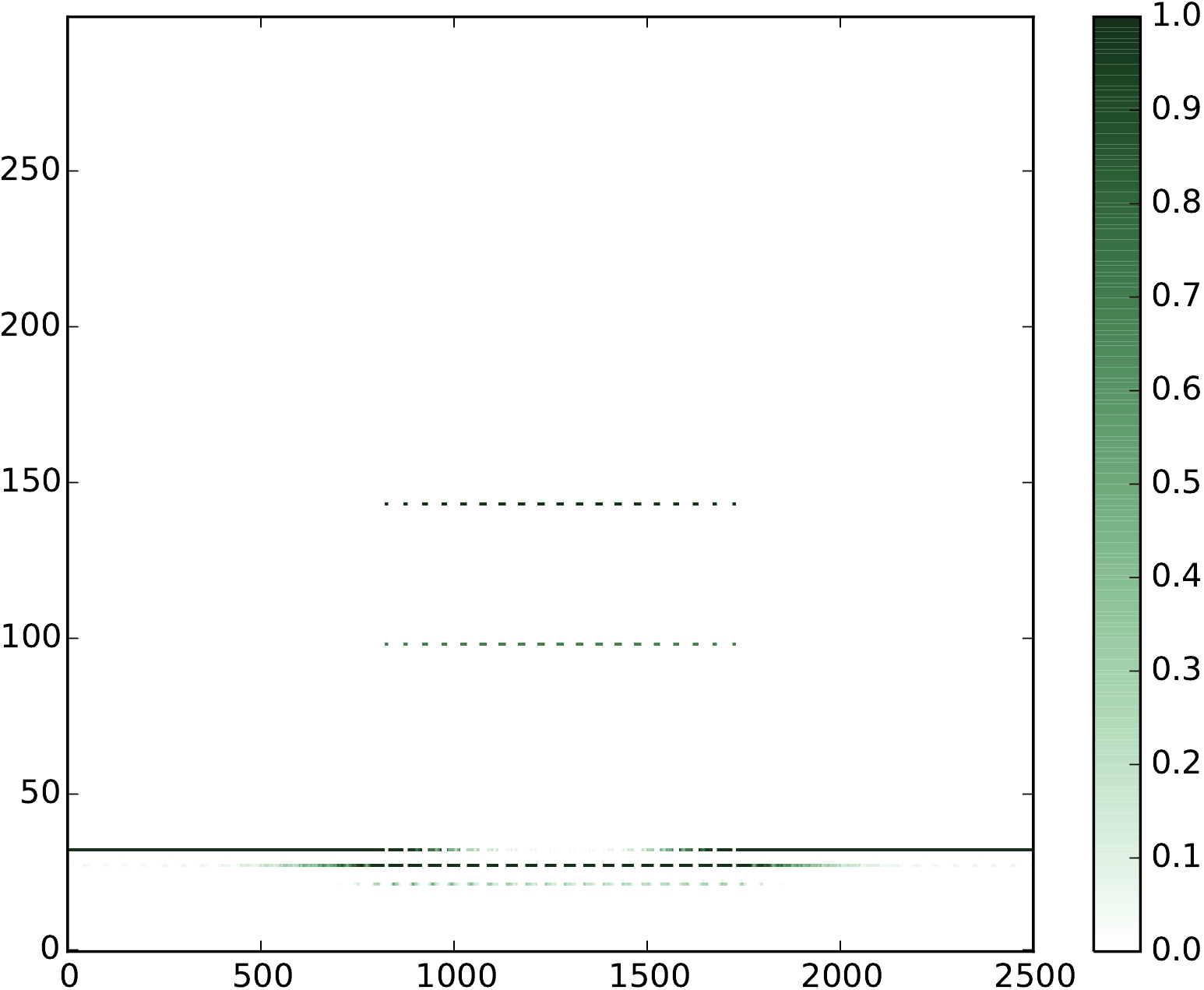}\\
    
                \includegraphics[width=.25\columnwidth]{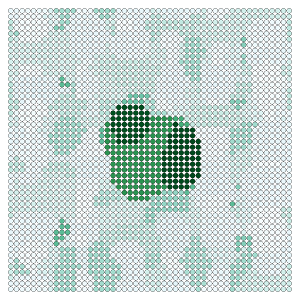}&
                \includegraphics[width=.25\columnwidth]{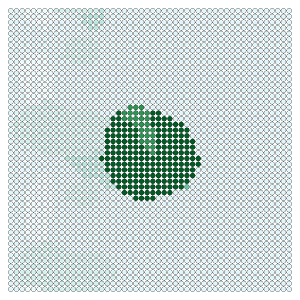}&
                \includegraphics[width=.25\columnwidth]{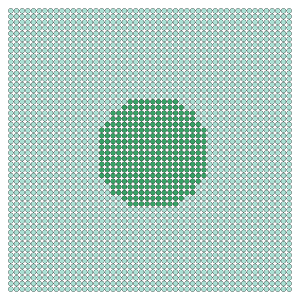}&
                \includegraphics[width=.25\columnwidth]{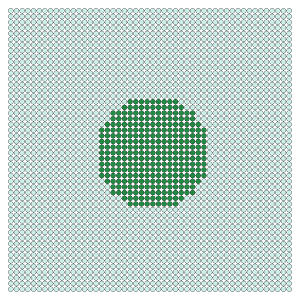}\\
            \end{tabular}
        \end{footnotesize}
        \setlength{\tabcolsep}{\tabcoldefault}
        
        \caption{
            \textbf{Top row}: Localized windows at the central vertex of the grid.
            \textbf{Center row}: Spectrograms.
            \textbf{Bottom row}: For each vertex, color represents the index of the frequency with maximum magnitude, from light green (low frequencies) to dark green (high frequencies).
        }
    \end{subfigure}
    
    \caption{All spectrograms coarsely identify the two sections in the input signal with different patterns (in all cases, we use the first $500$ eigenvectors only). In the unweighted case, the proposed method works significantly better than the convolutional approach~\cite{Shuman2012}. In the unweighted and weighted cases, the proposed PPR-based method has better spectral localization, i.e., for each vertex, fewer frequencies are selected.
    }
    \label{fig:grid}
\end{figure}

In the second example, we present results on a 2D grid graph, see \cref{fig:grid}. When this graph is unweighted, the graph Fourier transform amounts to the classical 2D Fourier. In the unweighted and weighted cases, the proposed short-graph Fourier transform is able to clearly identify the two different signal regions. Naturally, since the weight discontinuity matches the boundary between both signal regions, the spectrograms of the weighted graph have better spatial and frequency localizations. The proposed PPR-based spectrogram exhibits better spatial and frequency localizations than the convolutional approach.

For our last example, we use a real graph comprised of weather stations distributed throughout the US. The input signal is the average temperature in each station in 2014. The localized PPR windows follow nicely the graph topology, being more isotropic or anisotropic, depending on the local graph topology. The spectrogram obtained with the proposed PPR-based method presents clear patterns, which are coherent with the spatial arrangement of the graph vertices.

\begin{figure}[t]
    \centering
    
    \begin{subfigure}{\columnwidth}
        \centering
        \includegraphics[width=.93\columnwidth]{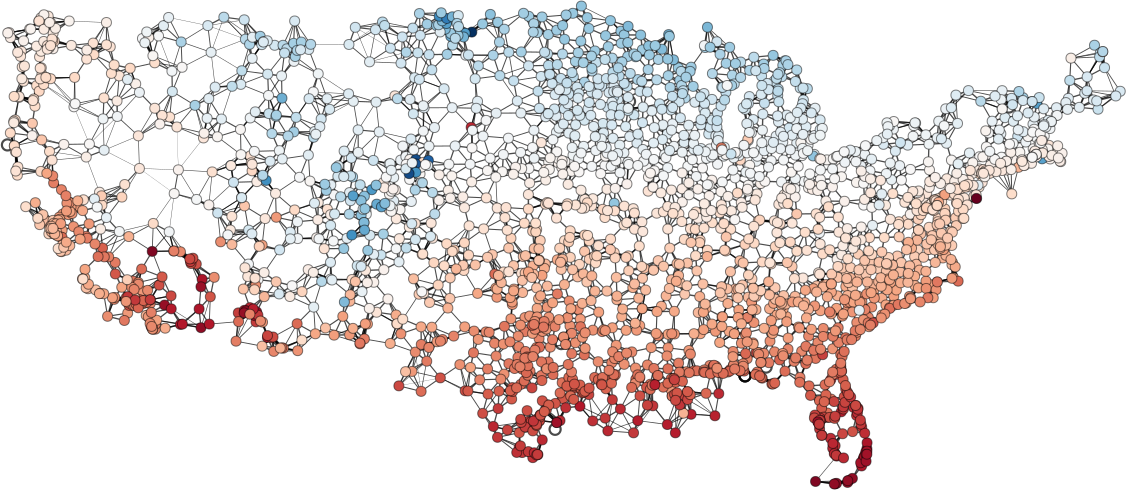}%
        \hfill%
        \includegraphics[width=.055\columnwidth]{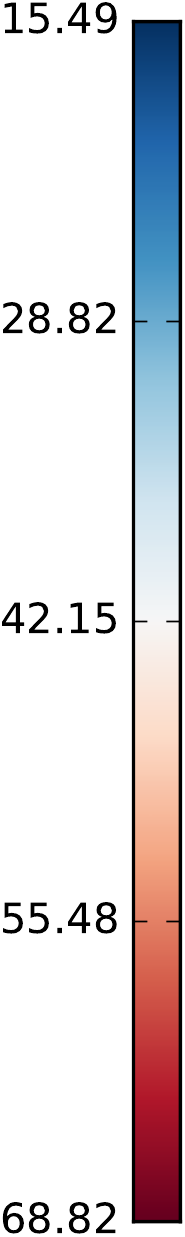}
        
        \caption{Graph of weather stations in the US, with color representing the average annual temperature in 2014. The graph was built by connecting each station to its 6 spatial nearest neighbors (the corresponding edge weight is the spatial distance between both stations).}
    \end{subfigure}

    \begin{subfigure}[b]{.35\columnwidth}
        \includegraphics[width=\columnwidth]{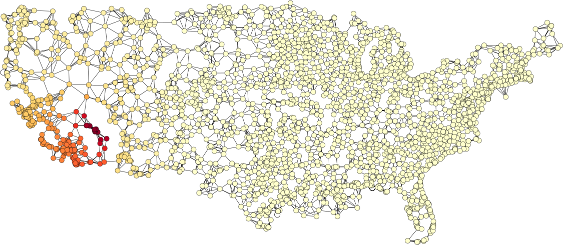}\\
        \includegraphics[width=\columnwidth]{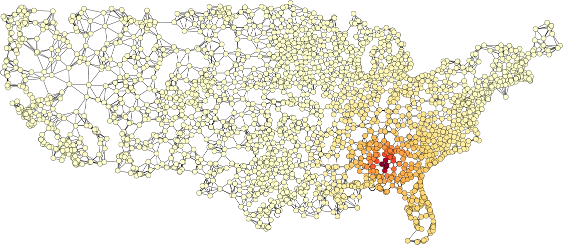}\\
        \includegraphics[width=\columnwidth]{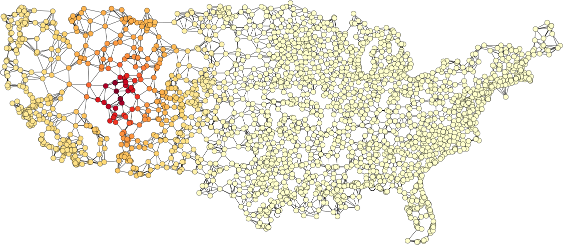}
        
        \caption{Different personalized PageRank windows.}
    \end{subfigure}%
    \hfill%
    \begin{subfigure}[b]{.63\columnwidth}
        \centering
        \includegraphics[width=.82\columnwidth]{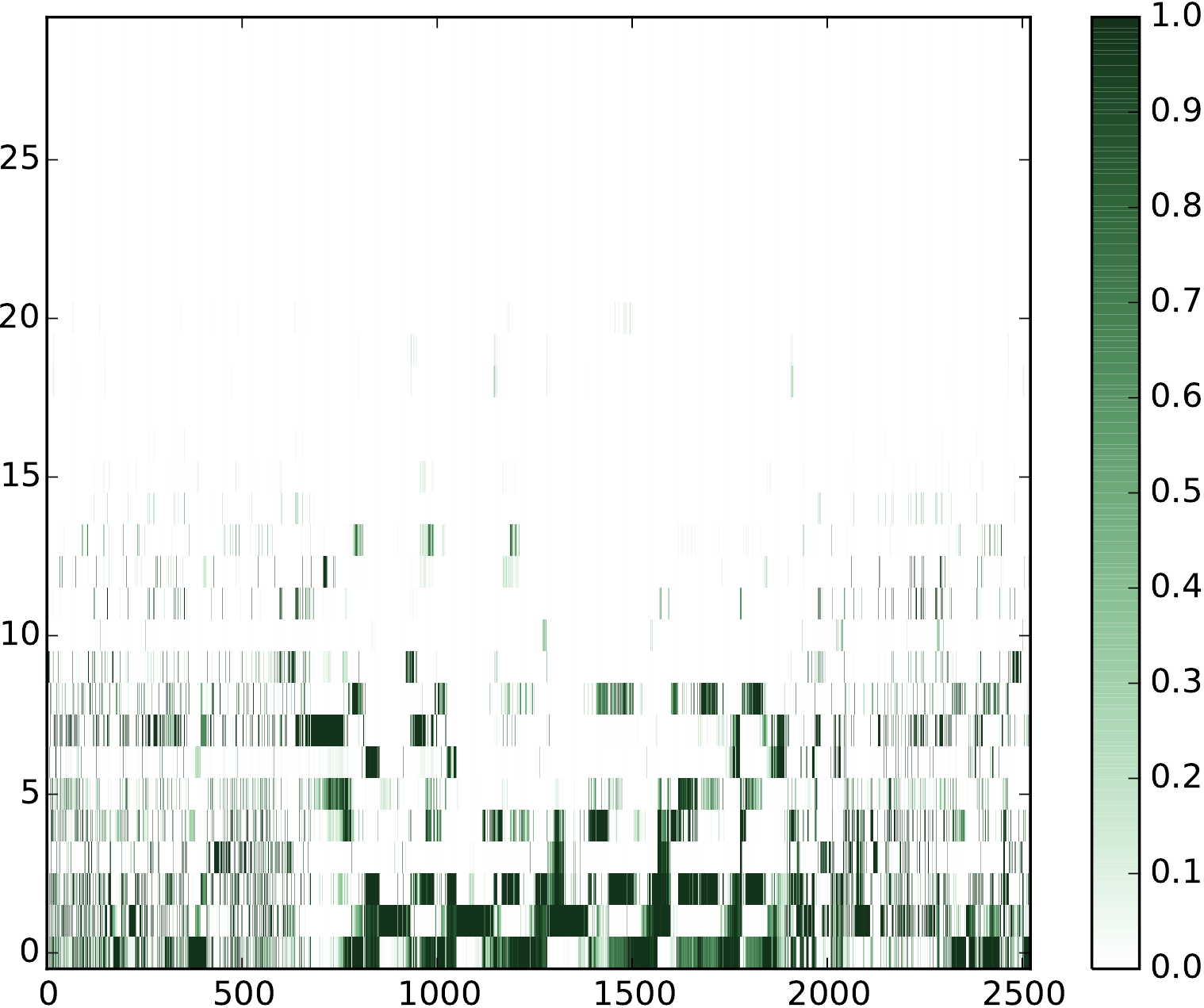}
        
        \caption{Spectrogram obtained with the proposed technique (for better visualization, we only show the first 30 components).}
    \end{subfigure}
    
    \begin{subfigure}{\columnwidth}
        \centering
        
    	\begin{tikzpicture}
        \node[anchor=south west,inner sep=0] (image) at (0,0) {
            \includegraphics[width=.93\columnwidth]{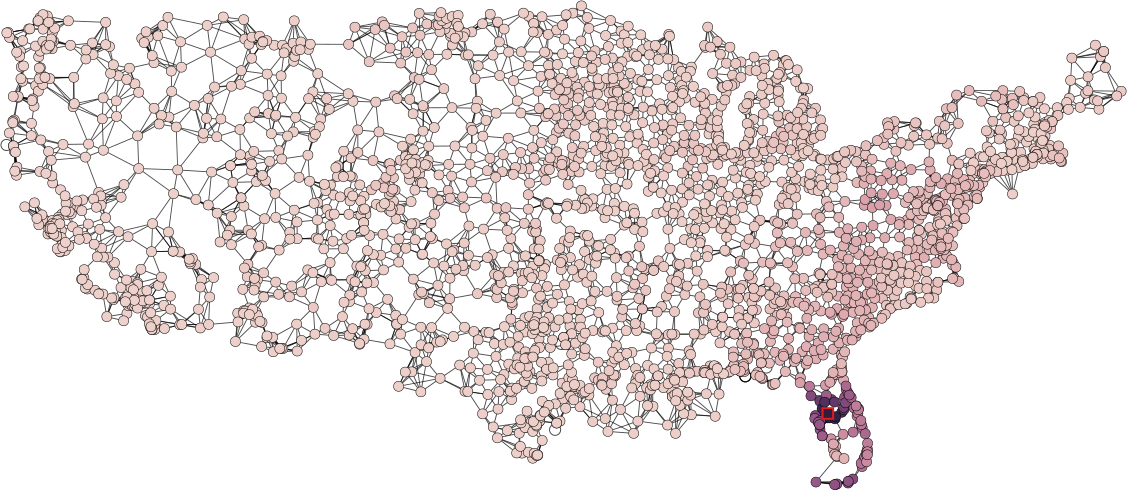}
        };
        \begin{scope}[x={(image.south east)},y={(image.north west)}]
            \draw [red, ->, very thick] (0.65, 0.10) -- (0.71, 0.14);
    	\end{scope}
        \end{tikzpicture}%
        \hfill%
        \includegraphics[width=.04\columnwidth]{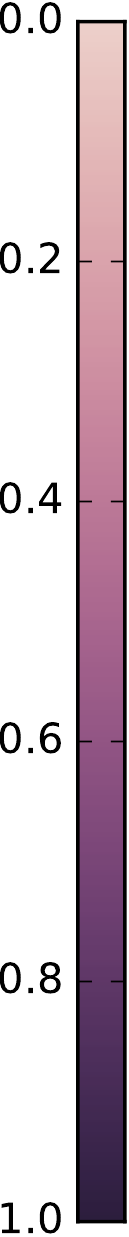}
        
        \caption{For the vertex marked with a red square (and a red arrow), we show the correlation between its spectral signature (its column in the spectrogram) and the signatures of the other vertices.}
    \end{subfigure}
    
    \caption{Clear patterns appear in the spectrogram, where nodes with similar spectral signature are localized in spatially coherent areas (e.g., Florida). We use $\beta = 10^{-3}$ and restrict the computations to the first $500$ eigenvectors.
    }
    \label{fig:temperature}
\end{figure}

\section{Conclusions}
\label{sec:conclusions}

In this work we presented an extension of the classical short-time Fourier transform (STFT) to signals defined over graphs.
We have shown with different examples that this new short-graph Fourier transform can be a valuable tool for extracting information from signals on graphs.

More broadly, we established the connection between local spectral graph theory and localized spectral analysis of graph signals. This is the first work that studies the use of personalized PageRank vectors as fundamental building blocks for local spectral analysis of graph signals.

The STFT becomes ineffective when the signal includes structures having different time-frequency resolution, some very localized in time and others in frequency. Wavelets address this issue by changing the time and frequency resolution. Such as diffusion wavelets extended the definition to the graph domain using powers of a graph diffusion operator, we plan on extending the use of personalized PageRank vectors to produce an alternate definition of wavelets on graphs.

\cleardoublepage

\bibliographystyle{IEEEbib}
\bibliography{sgft,graph_tool}

\end{document}